%% file: main.tex
\documentclass[11pt]{article}

\usepackage[preprint]{acl}

\usepackage{times}
\usepackage{latexsym}

\usepackage[T1]{fontenc}

\usepackage[utf8]{inputenc}

\usepackage{microtype}

\usepackage{inconsolata}

\usepackage{graphicx}

\usepackage{booktabs}   
\usepackage{natbib}     
\usepackage{diagbox}    
\usepackage{listings}
\usepackage{makecell}
\usepackage{tikz} 
\usepackage{shadowtext} 
\usepackage{colortbl}
\usepackage{xcolor} 
\definecolor{bluebackground}{HTML}{F0F8FF}  
\definecolor{blueframe}{HTML}{4682B4}       
\definecolor{greenbackground}{HTML}{E6F7E6}
\definecolor{greenframe}{HTML}{6BC16B}
\definecolor{shadowcolor}{RGB}{200,200,200} 
\newcommand{\boxshadow}[2][gray!40]{%
    \tikz[baseline=(text.base)]{%
        \node[inner sep=2pt, outer sep=0pt, fill=#1, minimum width=29pt, minimum height=13pt] (shadow) {}; 
        \node[inner sep=2pt, outer sep=0pt] (text) {#2}; 
    }%
}
\usepackage{placeins}
\usepackage{hyperref}
\usepackage{float}
\usepackage{tcolorbox} 
\tcbuselibrary{listings, breakable}
\usepackage{algorithm}
\usepackage{algorithmic}
\usepackage{amssymb}
\usepackage{booktabs}
\usepackage{pifont}
\usepackage{lipsum} 
\usepackage{float}
\usepackage{amsmath} 
\usepackage{graphicx}
\usepackage{multirow}
\usepackage{cite}

\usepackage{amsmath, amsthm}

\newtheorem{conclusion}{Conclusion}

\title{CoCo-Bench: A Comprehensive Code Benchmark For Multi-task Large Language Model Evaluation}

\author{
  \textbf{Wenjing Yin\textsuperscript{1}}\thanks{\texttt{ameliayin@stu.pku.edu.cn}},
  \textbf{Tianze Sun\textsuperscript{2}},
  \textbf{Yijiong Yu\textsuperscript{3}},
  \textbf{Jiawei Fang\textsuperscript{4}},
  \textbf{Guangyao Su\textsuperscript{4}},
  \\
  \textbf{Jiancheng Wang\textsuperscript{4}},
  \textbf{Zekun Wang\textsuperscript{5}},
  \textbf{Wei Wang\textsuperscript{5}},
  \textbf{Ran Chen\textsuperscript{5}},
  \textbf{Ziyun Dai\textsuperscript{5}},
  \\
  \textbf{Shuai Yuan\textsuperscript{1}},
  \textbf{Menghang Dong\textsuperscript{1}},
  \textbf{Peng Luo\textsuperscript{5}},
  \textbf{Dong Cao\textsuperscript{5}},
  \textbf{Da Lei\textsuperscript{5}},
  \textbf{Yajun Zhang\textsuperscript{5}},
  \\
  \textbf{Hao Chen\textsuperscript{5}},
  \textbf{Xiang Ma\textsuperscript{5}},
  \textbf{Yong Liu\textsuperscript{5}},
  \textbf{Weifeng Liu\textsuperscript{5}},
  \textbf{Yuanjian Xu\textsuperscript{6}},
  \textbf{Ji Pei\textsuperscript{5}},
\\
  \textsuperscript{1}Peking University,
  \textsuperscript{2}Harbin Institute of Technology,
  \textsuperscript{3}Tsinghua University,
  \\
  \textsuperscript{4}China Unicom Software Research Institute,
  \textsuperscript{5}OpenCSG,
  \\
  \textsuperscript{6}Hong Kong University of Science and Technology (Guangzhou)
}

\begin{document}
\maketitle
\input{src/0_abstract}
\input{src/1_introduction}
\input{src/2_related_work}
\input{src/3_task_definition}
\input{src/4_analysis}

\input{src/appendix/8_limitations_and_future_work}
\nocite{*} 
\bibliography{src/6_references} 
\input{src/8_appendix}
\end{document}

%% file: src/0_abstract.tex
\begin{abstract} 
Large language models (LLMs) play a crucial role in software engineering, excelling in tasks like code generation and maintenance. However, existing benchmarks are often narrow in scope, focusing on a specific task and lack a comprehensive evaluation framework that reflects real-world applications. To address these gaps, we introduce CoCo-Bench (Comprehensive Code Benchmark), designed to evaluate LLMs across four critical dimensions: code understanding, code generation, code modification, and code review. These dimensions capture essential developer needs, ensuring a more systematic and representative evaluation. CoCo-Bench includes multiple programming languages and varying task difficulties, with rigorous manual review to ensure data quality and accuracy. Empirical results show that CoCo-Bench aligns with existing benchmarks while uncovering significant variations in model performance, effectively highlighting strengths and weaknesses. By offering a holistic and objective evaluation, CoCo-Bench provides valuable insights to guide future research and technological advancements in code-oriented LLMs, establishing a reliable benchmark for the field.

\end{abstract}

%% file: src/1_introduction.tex
\section{Introduction}
In recent years, the application of artificial intelligence in software engineering (AI4SE)~\citep{mcdermott2020ai4se} has rapidly evolved, from code generation and bug detection to software testing. However, these advancements have exposed the limitations of existing code benchmarks. Many benchmarks fail to comprehensively assess large language models (LLMs), often overestimating their true performance and leading to biased conclusions.

\begin{figure*}[htbp]
    \centering
      \includegraphics[width=1\textwidth]{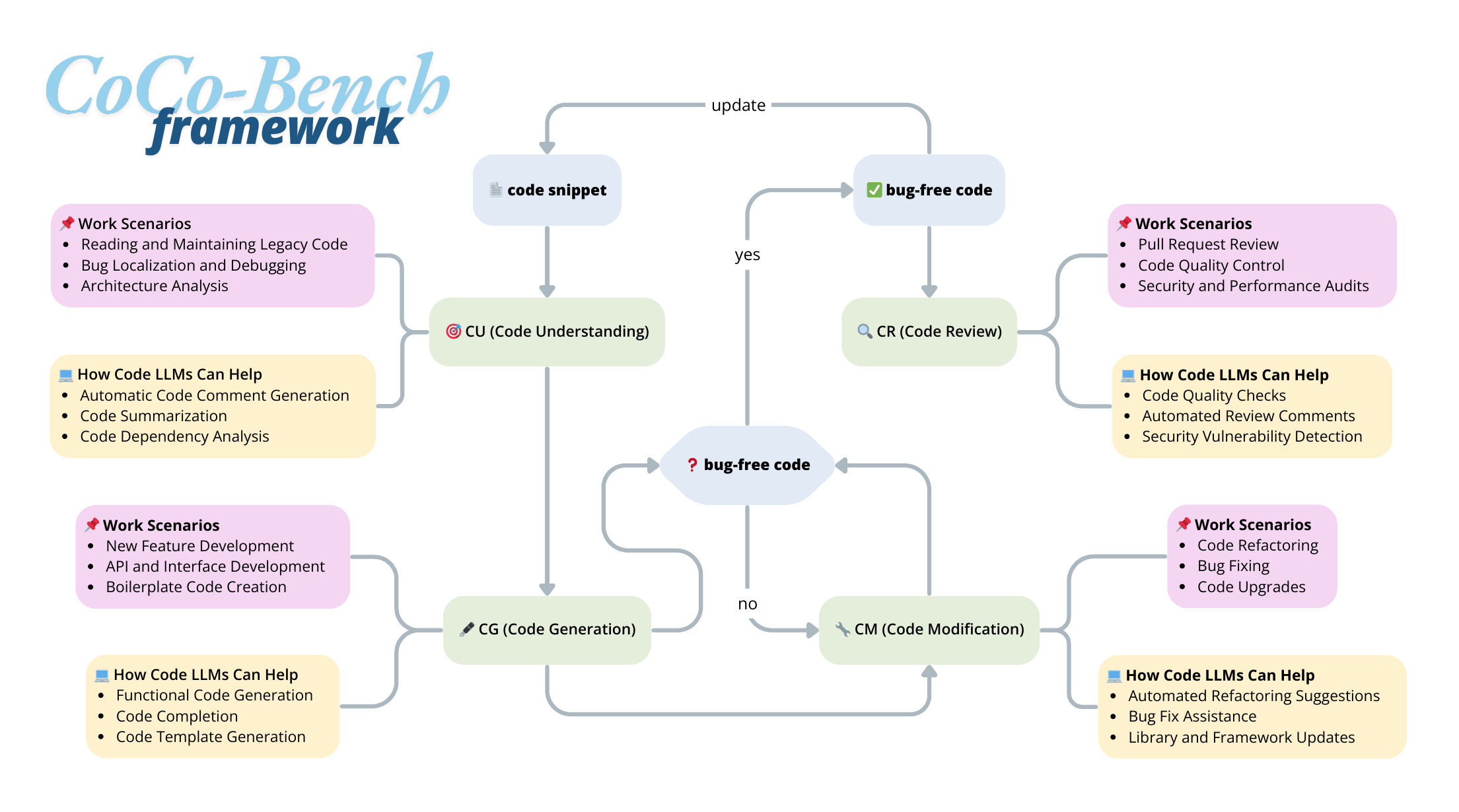}
    \caption{Overview of the core evaluation dimensions in CoCo-Bench. The framework assesses four critical capabilities of code LLMs: code understanding (CU), code generation (CG), code modification (CM), and code review (CR). The evaluation flow highlights the interconnected nature of these capabilities in real-world software development scenarios.}
    \label{fig:intro_fig}
\end{figure*}
As shown in \autoref{tab:tasks}, current benchmarks like HumanEval~\citep{chen2021evaluatinglargelanguagemodels} and MBPP~\citep{austin2021programsynthesislargelanguage} focus on simple test cases readily available online, making them prone to overfitting by models. Benchmarks such as CoderEval~\citep{Zhang_2024} and ClassEval~\citep{du2023classevalmanuallycraftedbenchmarkevaluating} target specific LLM capabilities but lack comprehensive coverage for a holistic evaluation. There is an urgent need for an objective, systematic, and comprehensive benchmark to support the continued development of AI4SE.

We propose that an effective benchmark should align closely with real-world scenarios and evaluate LLMs across four key dimensions of programmer capabilities: code understanding (CU, the ability to comprehend existing code), code generation (CG, the ability to generate code based on given context), code modification (CM, the ability to detect error and modify code), and code review (CR, the ability to assess and improve code quality). These four dimensions provide a robust framework for evaluating LLMs, ensuring a detailed and systematic measurement of their performance in practical development contexts.

To address these needs, we introduce CoCo-Bench, a Comprehensive Code Benchmark designed to assess LLMs across four key dimensions. It evaluates a range of programming languages and tasks of varying difficulty, with rigorous manual review to ensure quality and accuracy. Unlike existing benchmarks, CoCo-Bench features innovative task designs, such as reverse reasoning for CU and multi-level code completion for CG, offering a more comprehensive evaluation of model capabilities. CoCo-Bench ensures task diversity and practical alignment by including both simple and complex tasks. This approach prevents overfitting and better reflects model performance in real-world development environments, providing a nuanced framework that identifies strengths and weaknesses across different capability dimensions.

Our empirical results demonstrate that CoCo-Bench not only aligns well with existing benchmarks but also reveals significant variations in model performances across different capability dimensions. This effectively points out the strengths and weaknesses of various code LLMs. By offering a more holistic and objective evaluation, CoCo-Bench aims to guide future research, drive technological advancements in the development of code-oriented LLMs, and establish a reliable standard for the field of software engineering.

\begin{table}[htbp]
\centering
\begin{tabular}{lcccc}
\toprule
Code Benchmarks & CU & CG & CM & CR \\
\midrule
CoNaLA           & \ding{51} & \ding{55} & \ding{55} & \ding{55} \\
Concode          & \ding{51} & \ding{55} & \ding{55} & \ding{55} \\
HumanEval        & \ding{55} & \ding{51} & \ding{55} & \ding{55} \\
MBPP             & \ding{51} & \ding{55} & \ding{55} & \ding{55} \\
APPS             & \ding{51} & \ding{55} & \ding{55} & \ding{55} \\
PandasEval       & \ding{51} & \ding{55} & \ding{55} & \ding{55} \\
NumpyEval        & \ding{51} & \ding{55} & \ding{55} & \ding{55} \\
AixBench         & \ding{51} & \ding{55} & \ding{55} & \ding{55} \\
ClassEval        & \ding{51} & \ding{55} & \ding{55} & \ding{55} \\
CoderEval        & \ding{51} & \ding{55} & \ding{55} & \ding{55} \\
CodeFuseEval     & \ding{51} & \ding{51} & \ding{55} & \ding{55} \\
UltraEval        & \ding{55} & \ding{51} & \ding{55} & \ding{55} \\
CodeXGLUE        & \ding{51} & \ding{51} & \ding{51} & \ding{55} \\
NaturalCodeBench  & \ding{51} & \ding{55} & \ding{55} & \ding{55} \\
CodeScope        & \ding{51} & \ding{51} & \ding{51} & \ding{55} \\
Mercury          & \ding{51} & \ding{55} & \ding{55} & \ding{55} \\
ENAMEL           & \ding{51} & \ding{55} & \ding{55} & \ding{55} \\
\midrule
\textbf{CoCo-Bench} & \ding{51} & \ding{51} & \ding{51} & \ding{51} \\
\bottomrule
\end{tabular}
\caption{Task coverage across various code benchmarks. \ding{51} indicates coverage of certain type of task, while \ding{55} indicates no coverage.}
\label{tab:tasks}
\end{table}

%% file: src/2_related_work.tex
\section{Related Work}
Code benchmarks for LLMs have undergone significant evolution recently, reflecting notable advancements in the field. Early works like HumanEval~\citep{chen2021evaluatinglargelanguagemodels}, Mostly Basic Programming Problems (MPBB)~\citep{austin2021programsynthesislargelanguage} and The Code/Natural Language Challenge (CoNaLa)~\citep{yin2018learningalignedcodenatural} focused on fundamental CG tasks. HumanEval evaluated models’ ability to generate Python functions using real-world tasks, while MPBB included 974 tasks aimed at entry-level programmers, assessing CG from complex textual descriptions. 

Most code benchmarks primarily concentrate on tasks related to CG. APPS~\citep{hendrycks2021measuringcodingchallengecompetence} challenges models to generate Python code from natural language, simulating real-world developer tasks. CoderEval~\citep{Zhang_2024} expanded evaluations by focusing on non-standalone functions commonly found in open-source projects, providing a platform for assessing functional correctness. ClassEval~\citep{du2023classevalmanuallycraftedbenchmarkevaluating} introduced a benchmark for class-level CG, addressing gaps in existing evaluations by focusing on more complex tasks. Concode~\citep{iyer2018mappinglanguagecodeprogrammatic} targets generating Java class member functions from English documentation, addressing challenges in class member function generation. CodeXGLUE~\citep{lu2021codexgluemachinelearningbenchmark} and CodeEditorBench~\citep{guo2024codeeditorbenchevaluatingcodeediting} advance research in code understanding and editing, evaluating tasks like debugging and requirement switching. 

As task complexity increased, new benchmarks emerged to cover a broader range of scenarios. DyPyBench~\citep{Bouzenia_2024} is the first comprehensive benchmark for dynamic program analysis of Python projects. SWE-BENCH~\citep{jimenez2024swebenchlanguagemodelsresolve} evaluates models' ability to generate patches that pass real tests by linking GitHub issues with merged pull requests. CRUXEval~\citep{gu2024cruxevalbenchmarkcodereasoning} tests models on practical coding tasks using 800 different Python functions, while Debugbench~\citep{tian2024debugbenchevaluatingdebuggingcapability} focuses on assessing debugging capabilities, reflecting the need for more complex evaluations beyond CG. LiveCodeBench~\citep{jain2024livecodebenchholisticcontaminationfree} adopts a dynamic approach by continuously sourcing new programming challenges from competitive platforms to evaluate models' real-world capabilities, particularly in code self-repair and test output prediction. CodeMind~\citep{liu2024codemindframeworkchallengelarge} introduces dimensions like independent execution reasoning and specification reasoning to assess models' performance in complex tasks beyond simple CG. 

Multi-task benchmarks have also gained importance. CodeFuseEval~\citep{Di_2024} combines the standards of HumanEval-x and MPBB, introducing multi-task scenarios like code completion and cross-language translation. UltraEval~\citep{he2024ultraevallightweightplatformflexible} provides a lightweight, comprehensive framework to assess LLMs across various tasks, offering a unified evaluation platform. 

%% file: src/3_task_definition.tex
\section{Task Definition}

\begin{figure*}[h]
    \centering
    \includegraphics[width=\textwidth]{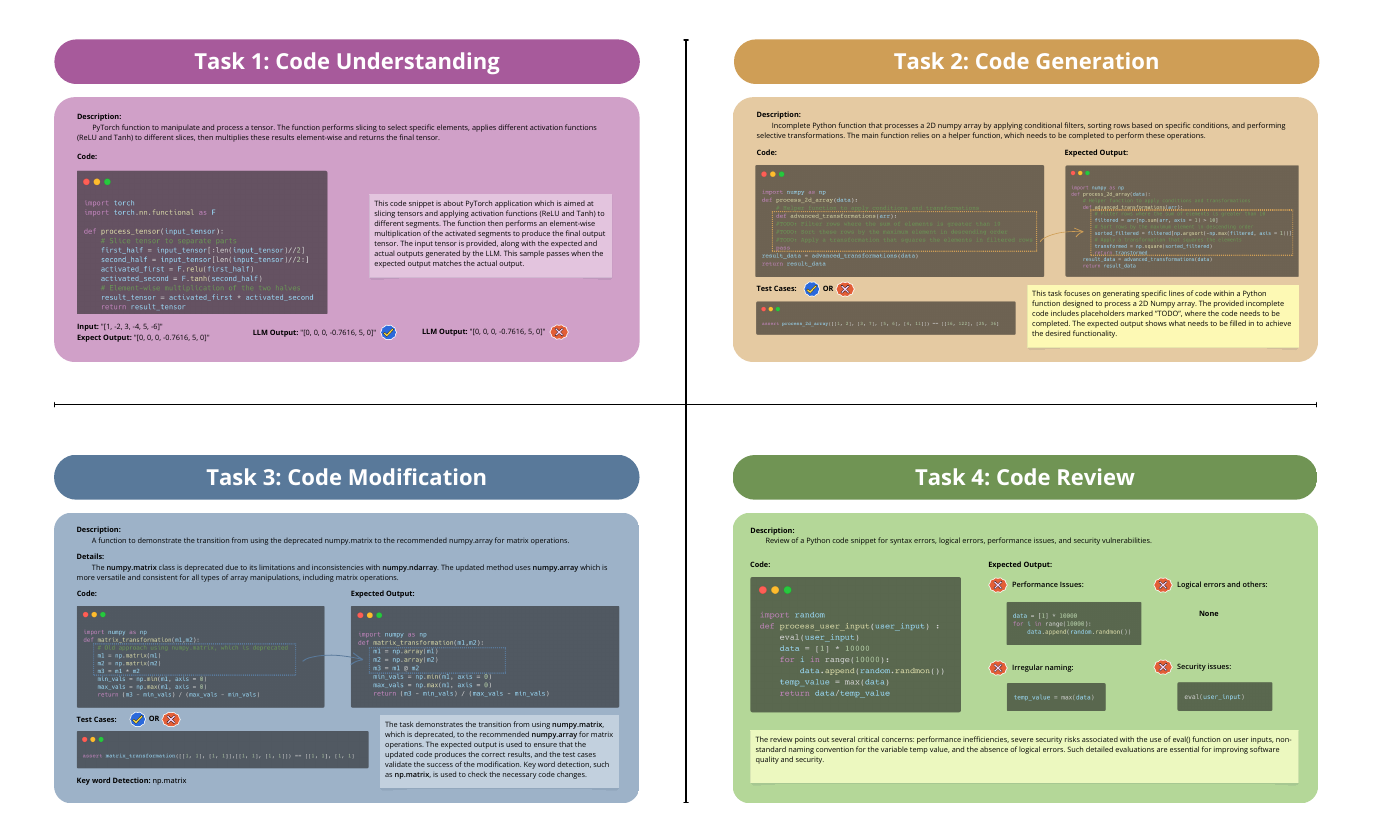}
    \caption{Illustration of the four primary tasks in CoCo-Bench—Code Understanding (CU), Code Generation (CG), Code Modification (CM), and Code Review (CR)—each defined to evaluate the capabilities of large language models (LLMs) in software engineering.}
    \label{fig:task_definition}
\end{figure*}

In CoCo-Bench, we define four primary tasks to comprehensively evaluate the capabilities of LLMs in software engineering. Each task is formally defined using mathematical notation to provide clarity and precision in assessment. Figure~\ref{fig:task_definition} shows some demonstrations.\\

\subsection{Code Understanding (CU)}
\label{sec:code_understanding}

CU is formalized as a bidirectional inference problem. Let \( \mathcal{C} \) denote the set of all possible code snippets, \( \mathcal{I} \) represent the set of all possible input parameters, and \( \mathcal{O} \) denote the set of all possible outputs. The CU task comprises two functions: \( f_{\text{CU}}: \mathcal{C} \times \mathcal{I} \rightarrow \mathcal{O} \) and \( f_{\text{CU}}^{-1}: \mathcal{C} \times \mathcal{O} \rightarrow \mathcal{P}(\mathcal{I}) \). Here, \( f_{\text{CU}}(C, I) = O \) predicts the output \( O \) given the code snippet \( C \) and input parameters \( I \), and \( f_{\text{CU}}^{-1}(C, O) = \mathcal{I}' \) deduces the set of possible input parameters \( \mathcal{I}' \subseteq \mathcal{I} \) that could produce the output \( O \) when executed with code \( C \).

Formally, for a given code snippet \( C \in \mathcal{C} \), input \( I \in \mathcal{I} \), and output \( O \in \mathcal{O} \), \( f_{\text{CU}}(C, I) = O \) if and only if \(\texttt{execute}(C, I) = O\), and \( f_{\text{CU}}^{-1}(C, O) = \{ I' \in \mathcal{I} \mid \texttt{execute}(C, I') = O \} \).

CU emphasizes the model's deep understanding of code logic and its ability to perform both forward and reverse inferences. To ensure wide applicability, the collected code snippets span multiple technical domains, including artificial intelligence and machine learning, data processing and analysis, web development, and database management. For example, in web development and data processing, we utilize key tools such as \texttt{PyTorch}~\citep{paszke2019pytorchimperativestylehighperformance}, \texttt{TensorFlow}~\citep{abadi2016tensorflowlargescalemachinelearning}, \texttt{Keras}~\citep{chollet2015}, \texttt{Scikit-learn}~\citep{pedregosa2018scikitlearnmachinelearningpython}, \texttt{NumPy}~\citep{harris2020array}, \texttt{Pandas}~\citep{mckinney-proc-scipy-2010}, and \texttt{Matplotlib}~\citep{Hunter:2007}.

Specifically, the CU task consists of two subtasks: predicting the code output (\( CU_F \)) and deducing the code input (\( CU_R \)). For instance, given the function \texttt{def add(a, b): return a + b} and inputs \( (3, 5) \), the model should accurately predict the result \( 8 \). Conversely, based on the output \( 8 \), the model should deduce potential inputs such as \( \{(3, 5), (4, 4), (6, 2)\} \). By assessing the model’s accuracy in both \( CU_F \) and \( CU_R \), we can comprehensively measure its understanding of code logic.

\subsection{Code Generation (CG)}
\label{sec:code_generation}

CG is defined as the transformation of natural language descriptions into executable code. Let \( \mathcal{D} \) represent the set of all possible natural language descriptions, and \( \mathcal{C} \) denote the set of all possible code snippets. The CG task is represented by the function \( f_{\text{CG}}: \mathcal{D} \rightarrow \mathcal{C} \), where \( f_{\text{CG}}(D) = C \) generates the code snippet \( C \) corresponding to the natural language description \( D \). Formally, for a given description \( D \in \mathcal{D} \), \( C = f_{\text{CG}}(D) \) such that \texttt{execute}(C) performs the task described by \( D \).

This function enables programmers to quickly translate ideas into code and allows non-technical individuals to contribute to software development. By automating routine programming tasks, CG reduces human errors, improves code quality and consistency, and accelerates the product iteration cycle. We design both function-level CG tasks and sentence-level CG tasks to test the models’ capabilities.

\subsection{Code Modification (CM)}
\label{sec:code_modification}

CM involves altering existing code to meet specific requirements or to improve its functionality. Let \( \mathcal{C}_{\text{old}} \) represent the set of original code snippets, \( \mathcal{R} \) denote the set of all possible modification requests or requirements, and \( \mathcal{C}_{\text{new}} \) denote the set of modified code snippets. The CM task is defined by the function \( f_{\text{CM}}: \mathcal{C}_{\text{old}} \times \mathcal{R} \rightarrow \mathcal{C}_{\text{new}} \), where \( f_{\text{CM}}(C_{\text{old}}, R) = C_{\text{new}} \) generates the modified code snippet \( C_{\text{new}} \) that satisfies the modification request \( R \). Formally, for given \( C_{\text{old}} \in \mathcal{C}_{\text{old}} \) and \( R \in \mathcal{R} \), \( C_{\text{new}} = f_{\text{CM}}(C_{\text{old}}, R) \) such that \(\texttt{execute}(C_{\text{new}})\) meets the requirements \( R \).

This task is crucial for maintaining software reliability and adapting to evolving needs. Our samples include code modification, where the model updates code based on error messages, and API updates, where the model revises code to reflect the latest API changes. These scenarios test whether code LLMs can enhance software reliability and keep applications up-to-date with the latest technological advancements.

\subsection{Code Review (CR)}
\label{sec:code_review}

CR is structured as a multi-label classification problem, where each code snippet is evaluated across several criteria. Let \( \mathcal{C} \) represent the set of all possible code snippets and \( \mathcal{E} \) denote the set of evaluation criteria. The CR task is defined by the function \( f_{\text{CR}}: \mathcal{C} \rightarrow \mathcal{P}(\mathcal{E}) \), where \( f_{\text{CR}}(C) = \mathcal{E}' \) assigns a subset of evaluation criteria \( \mathcal{E}' \subseteq \mathcal{E} \) that the code snippet \( C \) satisfies or violates. Formally, for a given code snippet \( C \in \mathcal{C} \), \( \mathcal{E}' = f_{\text{CR}}(C) \) where \( \mathcal{E}' = \{ e \in \mathcal{E} \mid C \text{ exhibits characteristic } e \} \).

The key areas of evaluation include security issues, performance problems, adherence to naming conventions, and logical errors. Specifically, security issues involve vulnerabilities such as SQL injections, buffer overflows, and insecure data handling. Performance problems pertain to code segments that may lead to inefficiency, including unnecessary computations, suboptimal algorithms, and resource-heavy operations. Adherence to naming conventions means the compliance with established coding standards and naming practices, which enhance code readability and maintainability. Logical errors involve flaws that might cause incorrect execution, such as infinite loops, improper condition checks, and erroneous data manipulation.

\begin{table*}[htbp]
    \centering
    \label{tab:model_comparison}
    \begin{tabular}{@{\hspace{1em}}l|cc|c|c|c|c@{\hspace{1em}}}
        \toprule
        \multirow{2}{*}{Model} & \multicolumn{2}{c|}{CU} & \multirow{2}{*}{CG} & \multirow{2}{*}{CM} & \multirow{2}{*}{CR} & \multirow{2}{*}{CoCo-Score}\\
        & $ \mathrm{CU_F} $ & $ \mathrm{CU_R} $ &  &  &  &\\
        \midrule
        CodeLlama-7b-base            & 12.93 & 3.33  & 11.45 & 30.00 & 39.39 & 16.87 \\
        CodeLlama-7b-instruct        & 9.92  & 5.17  & 16.41 & 15.00 & 25.71 & 12.94 \\
        CodeLlama-13b-base           & 19.01 & 4.17  & 19.23 & 20.00 & 32.26 & 16.67 \\
        CodeLlama-13b-instruct       & 13.45 & 6.78  & 18.60 & 25.00 & 31.43 & 17.14 \\
        CodeLlama-34b-base           & 15.83 & 6.67  & 21.71 & 15.00 & 20.00 & 14.50 \\
        CodeLlama-34b-instruct       & 14.88 & 4.24  & 19.08 & 20.00 & 29.41 & 15.48 \\
        \midrule
        DeepSeek-Coder-1.3b-base     & 14.88 & 3.31  & 15.27 & 20.00 & 31.43 & 14.84 \\
        Deepseek-Coder-1.3b-instruct & 15.13 & 5.98  & 19.85 & 20.00 & 25.71 & 15.62 \\
        DeepSeek-Coder-6.7b-base     & 26.05 & 5.08  & 25.78 & 25.00 & 28.57 & 19.68 \\
        Deepseek-Coder-6.7b-instruct & 35.65 & \textbf{11.93} & \textbf{46.92} & \underline{44.44} & 28.12 & 30.62 \\
        DeepSeek-Coder-33b-base      & 21.37 & 8.40  & 23.81 & 35.00 & 32.35 & 21.91 \\
        Deepseek-Coder-33b-instruct  & 33.88 & \underline{10.74} & 39.53 & \boxshadow{50.00} & \underline{37.14} & \underline{31.05} \\
        \midrule
        ChatGPT4                   & \underline{53.72} & \boxshadow{15.70} & \underline{44.27} & 15.00 & \boxshadow{45.71} & \textbf{32.06} \\
        DeepSeek-R1-Distill-Qwen-7B  & \textbf{57.02} & 9.09  & 35.88 & 20.00 & 34.29 & 28.26 \\
        o1-mini                    & \boxshadow{66.12} & 9.09 & \boxshadow{55.73} & \textbf{45.00} & \boxshadow{45.71} & \boxshadow{39.60} \\
        \bottomrule
    \end{tabular}
    \caption{Leaderboard of model performance comparison on CoCo-Bench, with the first-place models highlighted in \boxshadow{shadow}, the second-place models in \textbf{bold}, and the third-place models \underline{underlined}. The table compares the performance across different tasks: $\mathbf{CU}$, $\mathbf{CG}$, $\mathbf{CM}$, and $\mathbf{CR}$ for five major model series: DeepSeek-Coder, CodeLlama, R1, GPT and o1. $\mathbf{CU_F}$ and $\mathbf{CU_R}$ are two kinds of sub-tasks of $\mathbf{CU}$. The metric CoCo-Score (see ~\ref{app:metrics}) provides an aggregated evaluation of model capabilities.}
    \label{model_comparison}
\end{table*}

%% file: src/4_analysis.tex
\section{Analysis}

In this section, we first analyze the overall scores of all the LLMs. Then, in Section ~\ref{sec:corr}, we examine the correlations between different sub-tasks and datasets. In Section~\ref{sec:context}, we analyze the impact of context length on model inference. In Section~\ref{sec:decoding}, we focus on the impact of decoding strategies on the models, primarily including top-k, top-p, and Max New Tokens.

\subsection{Comprehensive Performance Evaluation}

We present the leaderboard results in Table \ref{model_comparison}, in which the empirical results reveal that DeepSeek-Coder-V2-Instruct and ChatGPT-4.0 outperform other leading LLMs across various code-related tasks. This superior performance primarily stems from the scaling up of model parameters and the volume of data used during their pre-training processes.

Model scaling plays a pivotal role in enhancing performance. Larger models, such as DeepSeek-Coder-V2-Instruct, consistently outperform smaller models like DeepSeek-Coder-1.3b and DeepSeek-Coder-6.7b. This scaling trend is not uniformly observed across all model families, suggesting that factors such as training data diversity and fine-tuning strategies play critical roles in maximizing the benefits of larger model architectures. For instance, while DeepSeek-Coder-33b-instruct achieves significant performance gains, similar scaling within the CodeLlama series does not yield comparable improvements, pointing to potential inefficiencies or bottlenecks in their respective training paradigms.

The superiority of instruction-tuned models over their base counterparts is consistently observed across both CoCo-Bench and HumanEval benchmarks. This trend underscores the critical importance of instruction fine-tuning in enhancing models' ability to follow complex directives and perform specialized tasks. The enhanced performance of larger instruction-tuned models further suggests that combining scale with targeted fine-tuning yields synergistic benefits, enabling models to achieve higher levels of proficiency in intricate code-related tasks.

\subsection{Correlation Analysis}
\label{sec:corr}
\begin{figure}[H]
    \centering
    \includegraphics[width=0.8\linewidth]{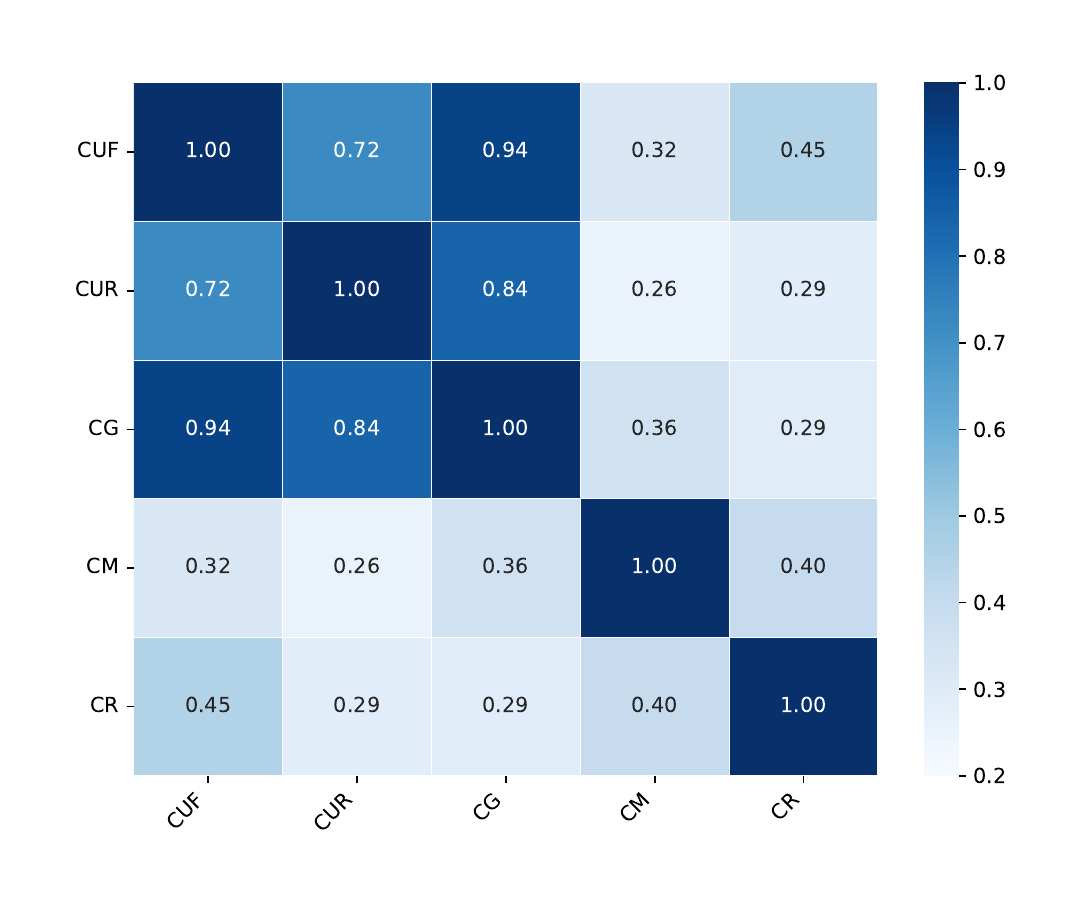}
    \caption{The correlation between each two of the five tasks (CU\textsubscript{F}, CU\textsubscript{R}, CG, CM, CR) on CoCo-Bench.}
    \label{fig:heatmap}
\end{figure}

\paragraph{Correlation between tasks:} We visualize the Spearman correlation heatmap among the tasks in \autoref{fig:heatmap}, which provides further insight into the interdependencies between different code-related tasks. Notably, a strong correlation is observed between \(\text{CU}_F\) and CG (0.94), as well as between \(\text{CU}_R\) and CG (0.84). These high correlation values indicate that proficiency in code understanding (CU) significantly enhances a model's ability to generate (CG) code effectively. This interconnectedness suggests that improvements in one task—such as CU—can lead to better performance in related tasks like CG. It highlights the integrated nature of these tasks in real-world applications, where advances in one area may catalyze progress in others, ultimately fostering more versatile and capable code models.

However, the relatively lower correlation between CU and CR indicates that CR involves additional competencies beyond mere code comprehension, such as assessing code quality, efficiency, and adherence to best practices. We also observe a relatively low correlation between CG and CM (0.36), suggesting that the performance in CG does not necessarily contribute to CM. This distinction underscores the necessity for models to possess not only a deep understanding of code but also the capability to evaluate and improve it critically, reflecting the multifaceted nature of software engineering.

\begin{figure}[htbp]
    \centering
    \includegraphics[width=1.0\linewidth]{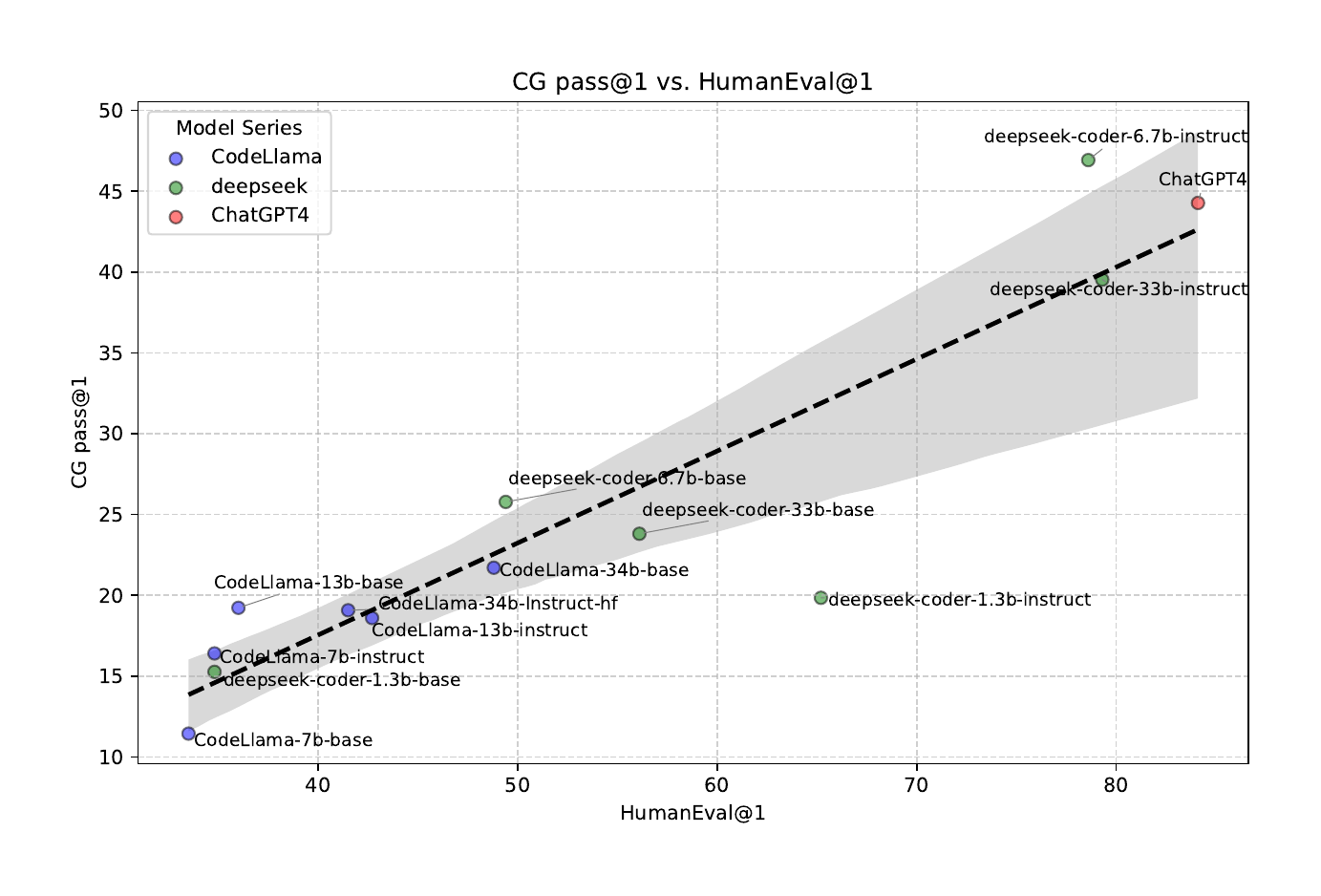}
    \caption{Comparative analysis of model performance on CoCo-Bench and HumanEval benchmarks: This figure illustrates the relationship between model performance on the CoCo-Bench (CG pass@1) and HumanEval (HumanEval@1) benchmarks. The dashed trend line and shaded area indicate the general correlation between performance on the two benchmarks.}
    \label{cgvshumaneval}
    \end{figure}

\paragraph{Correlation between datasets:} \autoref{fig:heatmap} and \autoref{cgvshumaneval} collectively demonstrate that CoCo-Bench provides a more stringent and discriminative evaluation compared to simpler benchmarks like HumanEval. The positive correlation between CoCo-Bench and HumanEval scores indicates that foundational capabilities are consistent across benchmarks. Moreover, the analysis reveals that increasing model size generally correlates with enhanced performance, particularly within the DeepSeek-Coder series. 

\begin{figure}[H]
    \centering
    \includegraphics[width=1.0\linewidth]{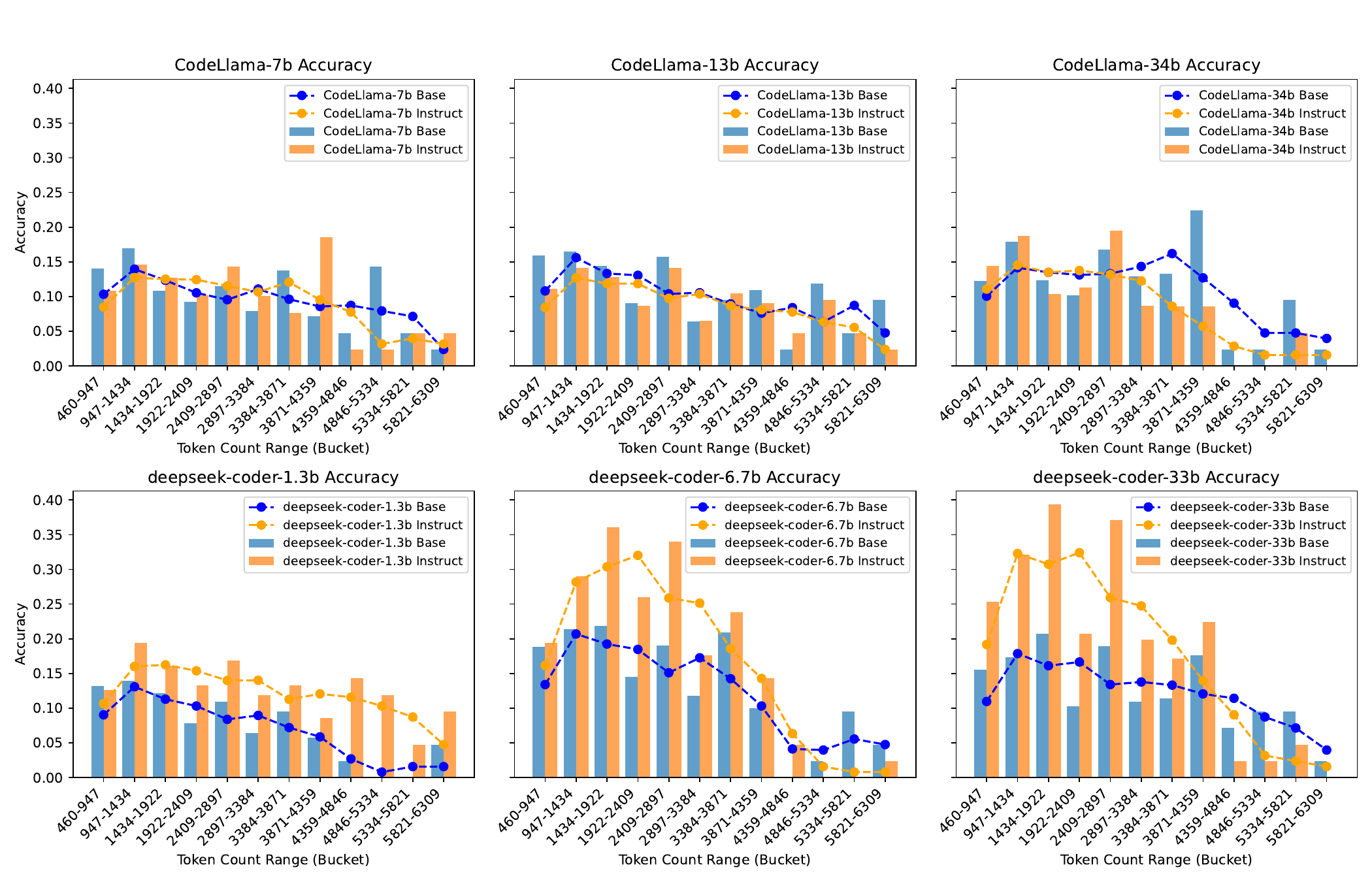}
    \caption{Performance differences between instruct and base versions of the same models on CoCo-Bench}
    \label{fig:basevsinstruct2}
\end{figure}

\begin{figure*}[htbp]
  \centering
  \includegraphics[width=1\linewidth]{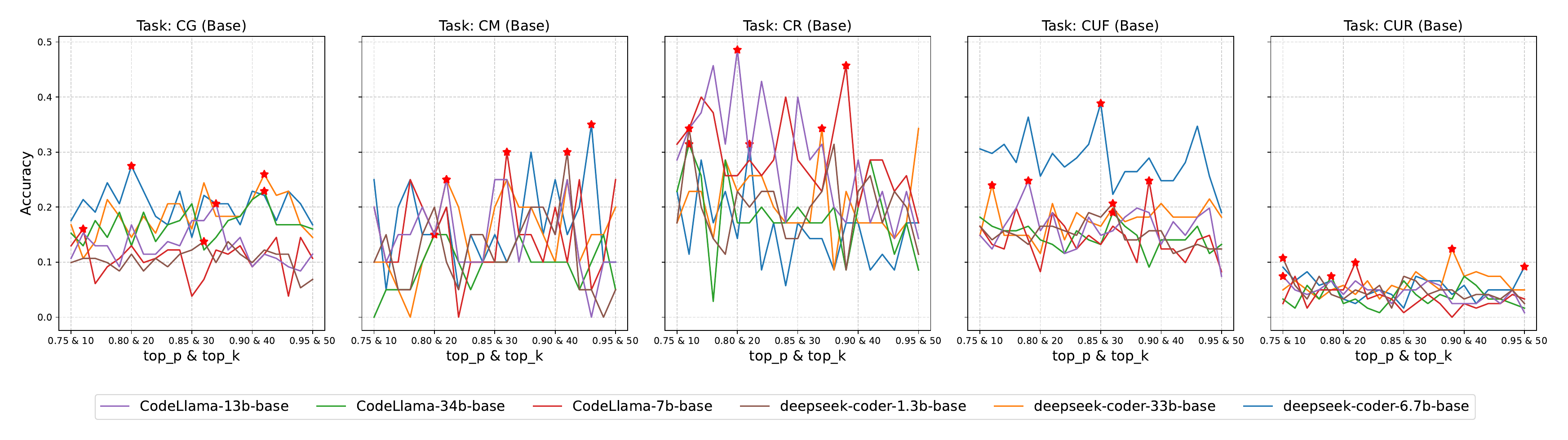}
  \\ 
  \vspace{0cm} 
  \includegraphics[width=1\linewidth]{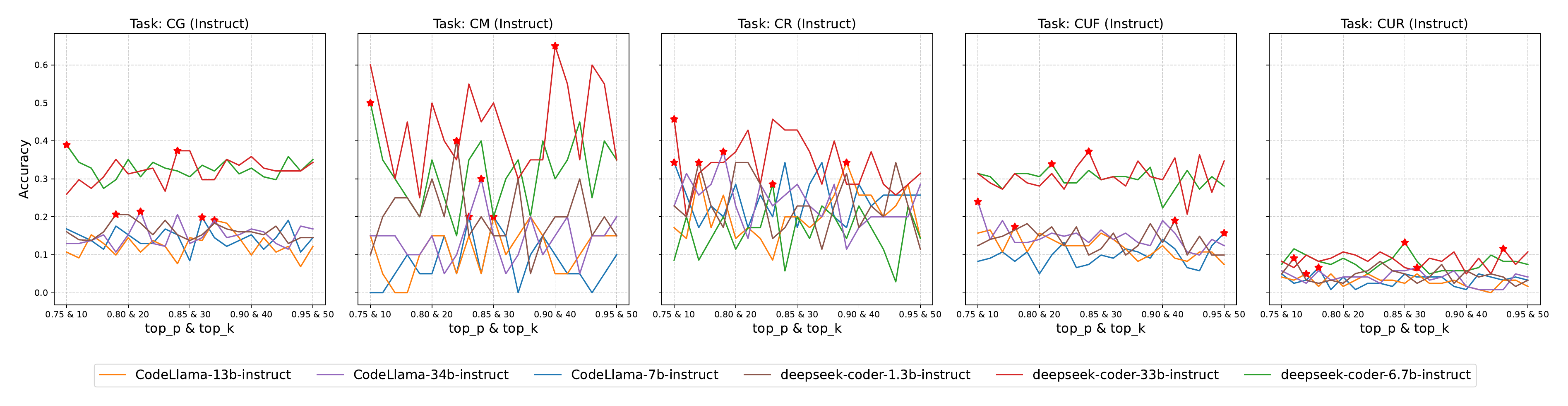}
  \caption{Performance differences under various \texttt{top\_p} and \texttt{top\_k} configurations}
  \label{fig:topp&topk}
\end{figure*}

\subsection{Context Length and Model Performance}
\label{sec:context}
\begin{figure}[H]
    \centering
    \includegraphics[width=1.0\linewidth]{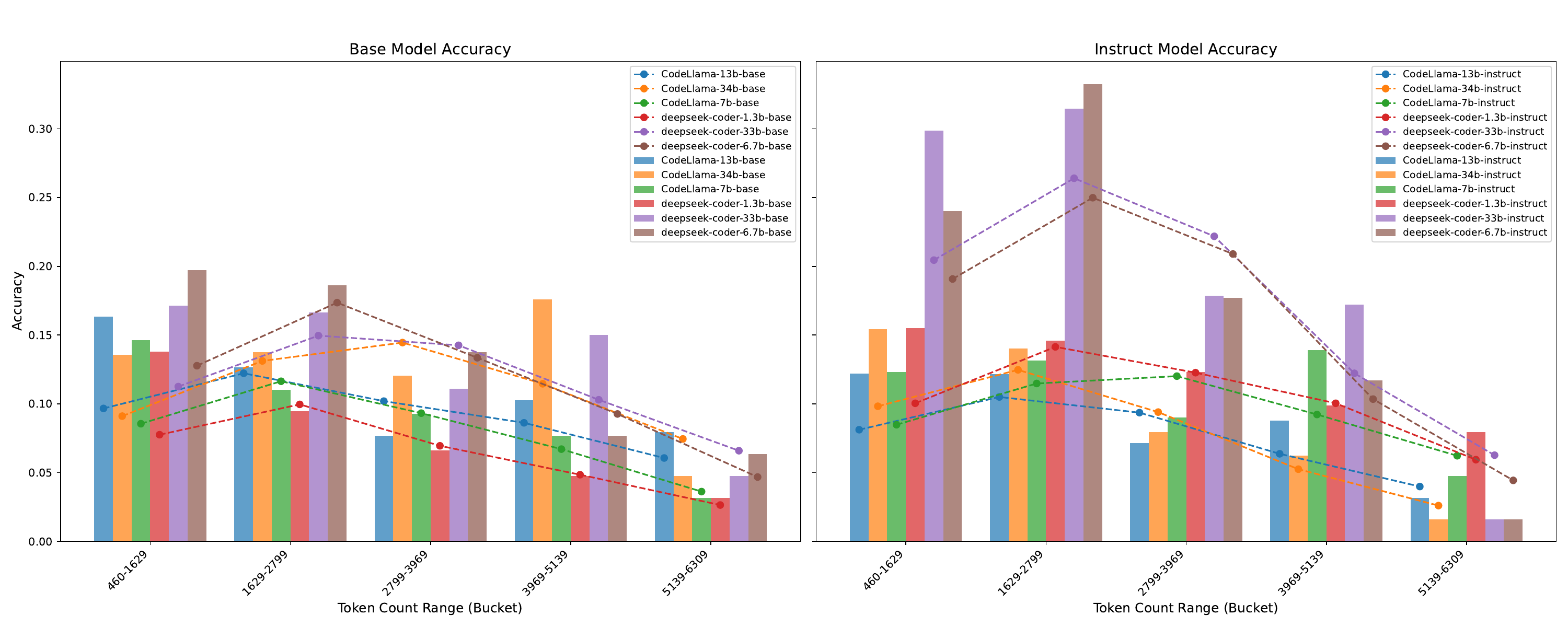}
    \caption{Performance differences between different models of the same type on CoCo-Bench}
    \label{fig:basevsinstruct1}
\end{figure}

 \autoref{fig:basevsinstruct2} and \autoref{fig:basevsinstruct1} illustrate how model accuracy varies across different token count ranges of input. All models, regardless of size and type, face challenges in maintaining attention across longer contexts due to the quadratic cost of self-attention in transformers. Instruction tuning likely helps focus attention on task-relevant segments of the input, mitigating some of this drop in performance for instruction models. 

\subsection{Decoding Strategies}
\label{sec:decoding}
\begin{figure*}[htbp]
  \centering
  \includegraphics[width=1\linewidth]{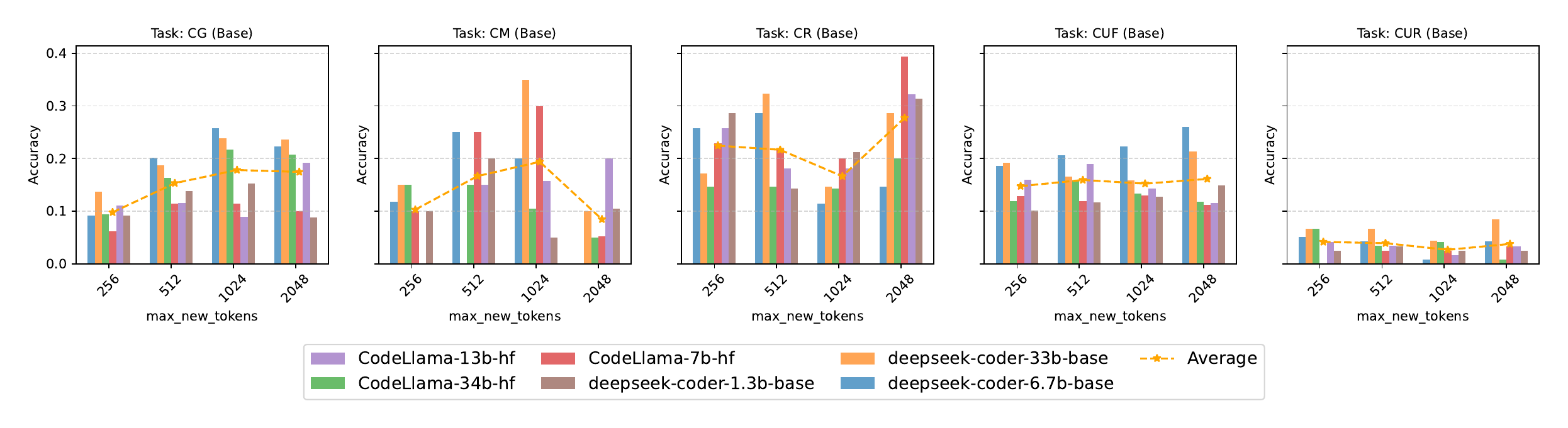}
  \\ 
  \vspace{0cm} 
  \includegraphics[width=1\linewidth]{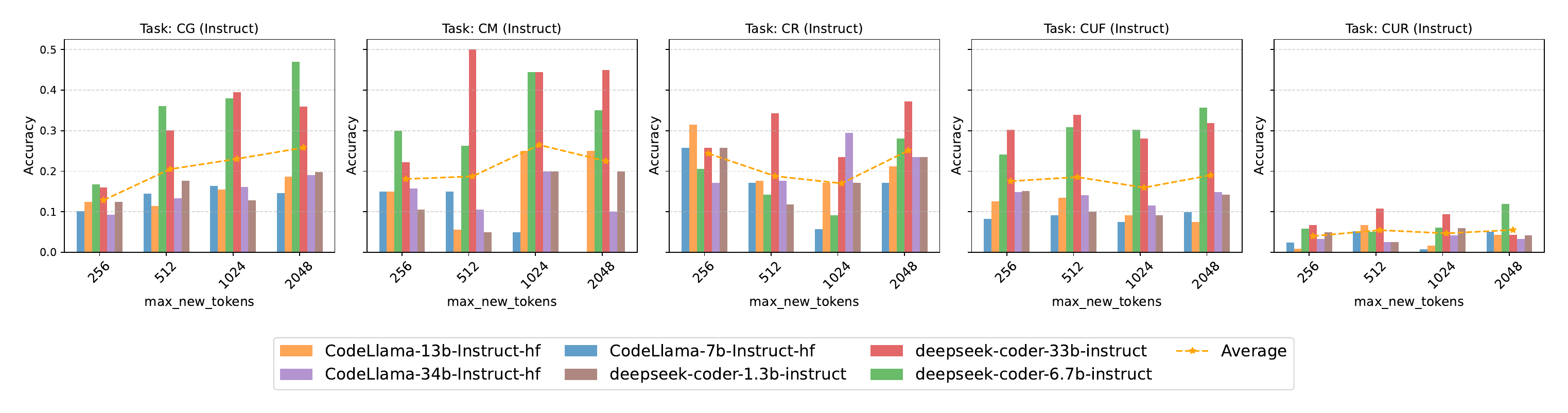}
  \caption{Performance differences under various \texttt{max\_new\_tokens} Configurations}
  \label{fig:maxnewtokens}
\end{figure*}
\paragraph{Top-p and Top-k:} As is illustrated in \autoref{fig:topp&topk}, top-p and top-k sensitivity varies by task. Structured tasks like CU tasks benefit from deterministic outputs since code has stricter correctness requirements. A narrower token distribution (lower top-p and top-k) is often sufficient for accurate results. In contrast, open-ended tasks like CM and CG require creativity and diversity from LLMs, as there are multiple plausible ways to modify or improve code. It's worth mentioning that LLMs are more sensitive to top-p and top-k variance as they generate a wider range of plausible token predictions due to their richer token distribution and greater expressive power. When using high top-p and top-k for inference, the outputs of LLMs can become overly diverse or less coherent, potentially introducing syntactic errors or irrelevant code snippets. Smaller models, by contrast, have less expressive token distributions, making them inherently more deterministic and less impacted by high top-p and top-k settings, thereby maintaining code correctness more effectively. So, we can draw the following empirical conclusion:
\begin{conclusion}
The sensitivity to top-p and top-k parameters varies across different tasks. Structured tasks, such as CU, tend to benefit from more deterministic outputs, while open-ended tasks, such as CM and CG, demand greater creativity and diversity.
\end{conclusion}

Instruction models consistently outperform base models in code-related tasks. We believe this is due to the exposure of instruction models to examples with task-specific instructions during training, enabling them to better understand and follow coding guidelines and requirements. As a result, we can derive the following empirical conclusion:
\begin{conclusion}
    Instruction-tuned models excel in code-related tasks due to their close alignment with task-specific coding objectives, which significantly enhances both performance and robustness across various decoding configurations.
\end{conclusion}

Medium token ranges align well with the pretraining datasets and tokenization schemes for most models. Shorter tokens (e.g., 460–1191) may lead to sparse representation, while longer tokens may overload the model's capacity. It's also noted how scaling laws reconcile with the outcome. Larger models are inherently better at capturing long-range dependencies due to their larger parameter space and richer latent representations. Smaller models lack the capacity to encode such complexity effectively. This leads us to the following empirical conclusion:

\begin{conclusion}
    LLMs encounter challenges with longer code contexts, but instruction tuning helps maintain performance by focusing on relevant code segments. Larger models handle long-range dependencies in code more effectively than smaller ones.
\end{conclusion}

\textbf{Max New Tokens:} \autoref{fig:maxnewtokens} shows that higher max new tokens directly correlates with better accuracy in tasks requiring coherent and extended outputs (e.g., CG and CU). It allows the model to generate longer sequences, potentially capturing more context and completing more complex outputs. Some tasks like CM benefit less due to their inherent requirements, where longer sequences add minimal value. Comparably, larger models and instruction-tuned models are better equipped to utilize extended token generation due to their ability to manage more extensive contexts and dependencies. Based on this, the following empirical conclusion can be drawn:

\begin{conclusion}
    Increasing max new tokens improves performance in code tasks that require extended and coherent outputs, particularly benefiting larger and instruction-tuned models by enabling them to manage more complex code structures and dependencies.
\end{conclusion}

%% file: src/appendix/8_limitations_and_future_work.tex
\section{Limitations and Future Work}

While CoCo-Bench provides an effective and comprehensive evaluation framework for LLMs in code-related tasks, there are areas for expansion. Currently, the benchmark lacks multimodal tasks that require models to integrate code with other data types, such as images or natural language, which are becoming increasingly relevant in modern development environments. Recognizing this gap, we plan to introduce multimodal tasks in the future, allowing us to evaluate models on  more complex projects and providing a more holistic assessment of LLMs' capabilities. To ensure CoCo-Bench remains at the cutting edge of LLM evaluation, we will regularly update the benchmark by incorporating new programming languages and adapting to evolving development practices.

%% file: src/8_appendix.tex
\newpage
\appendix
\input{src/appendix/1_data_statistics_details}

\input{src/appendix/3_benchmark_construction}
\input{src/appendix/4_benchmark_metric}
\input{src/appendix/5_prompts_for_inference}
\input{src/appendix/6_more_hyper_parameter_analysis}

\newpage

%% file: src/appendix/1_data_statistics_details.tex
\section{Data Statistics Details}
\subsection{Data Statistics}
\begin{figure}[htbp]
    \centering
    \includegraphics[width=1.0\linewidth]{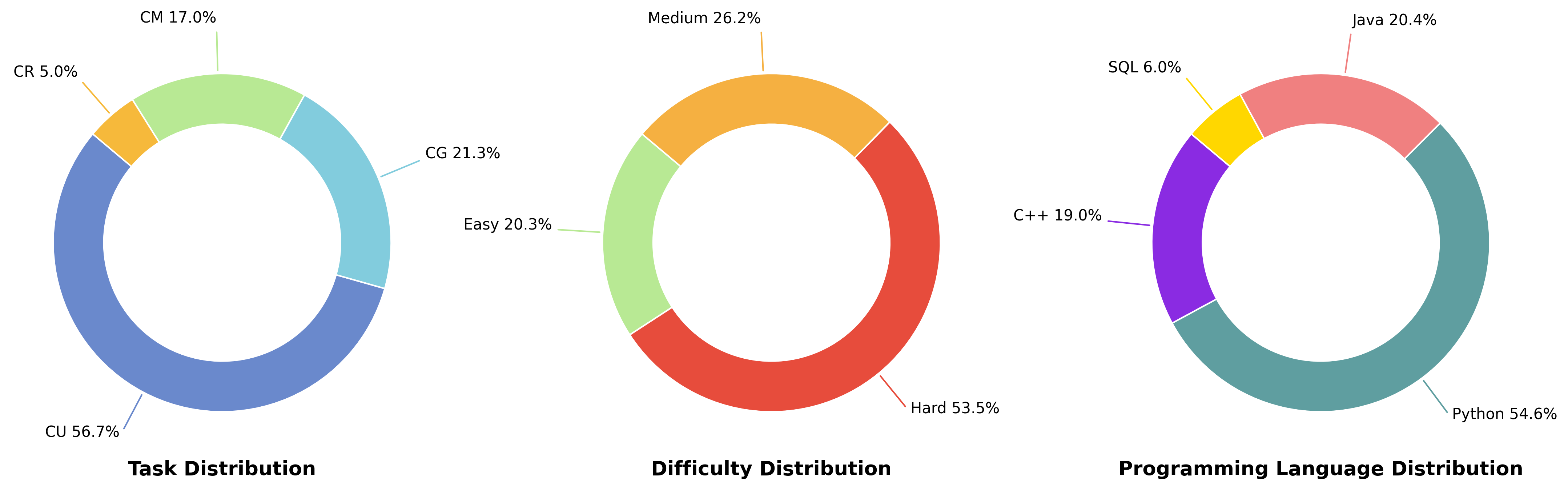}
    \caption{Distribution of tasks, difficulty levels, and programming languages in CoCo-Bench. }
    \label{fig:enter-label}
\end{figure}

CoCo-Bench, meticulously curated through fine-grained manual review, comprises 705 high-quality samples developed with the assistance of several seasoned developers, each with over 10 years of experience in SE. Half of CoCo-Bench samples are quite challenging. As shown in Figure ~\ref{fig:enter-label}, The tasks are diversified, with 56.7\% of the samples focused on CU, 21.3\% on CG, 17.0\% on CM, and 5.0\% on CR. This distribution emphasizes a comprehensive approach to testing the various aspects of code comprehension and generation. 

In addition to the challenging nature of the tasks, CoCo-Bench also supports a wide range of programming languages. Python leads the way, accounting for 54.6\% of the samples, followed by Java at 20.4\%, C++ at 19.0\%, and SQL at 6.0\%. The diversity ensures that the models are evaluated across a broad spectrum of real-world coding scenarios.

\begin{table}[htbp]
\centering
\begin{tabular}{lccccc}
\toprule
 & \multicolumn{4}{c}{Task Types} &   \\
\cmidrule(lr){2-5}
Difficulty & CU & CG & CM & CR & Total \\
\hline
Easy       & 74 & 41 & 17 & 11 & 143 \\
Medium    & 94 & 54 & 22 & 15 & 185 \\
Hard   & 232 & 55 & 81 & 9  & 377 \\
\hline
Total                         & 400 & 150 & 120 & 35 & 705 \\
\bottomrule
\end{tabular}
\caption{Distribution of samples by difficulty levels. This table presents the distribution of samples across different difficulty levels in the dataset. The tasks are categorized into four types: CU, CG, CU and CR. Each sample is further divided into easy, medium and hard difficulty levels. }
\label{tab:difficulty}
\end{table}
\begin{table}[htbp]
\centering
\begin{tabular}{lccccc}
\toprule
 & \multicolumn{4}{c}{Task Types} &   \\
\cmidrule(lr){2-5}
Code Language & CU & CG & CM & CR & Total \\
\hline
python       & 208 & 87 & 55 & 35 & 385 \\
java    & 80 & 29 & 35 & \diagbox{}{}  & 144 \\
C++   & 70 & 34 & 30 & \diagbox{}{}   & 134 \\
SQL   & 42 & \diagbox{}{}  & \diagbox{}{}  & \diagbox{}{}   & 42 \\
\hline
Total                         & 400 & 150 & 120 & 35 & 705 \\
\bottomrule
\end{tabular}
\caption{Distribution of task types by programming languages. Each row represents a programming language, such as Python, Java, C++ and SQL, and lists the number of tasks in each category for that language. The total number of tasks for each language is provided, along with the overall total for all languages. }
\label{tab:language}
\end{table}

The dataset, as summarized in Table~\ref{tab:difficulty}, shows a clear emphasis on hard examples, which constitute the majority of the dataset with 377 tasks. Additionally, as shown in Table~\ref{tab:language}, each task type is well-represented across different languages, with Python being the most frequently used language across all categories.

%% file: src/appendix/3_benchmark_construction.tex
\section{Benchmark Construction}
\begin{figure*}[t]
\centering
\hspace*{-2.2cm}
\vspace*{-2cm}
\includegraphics[width=1.25\textwidth]{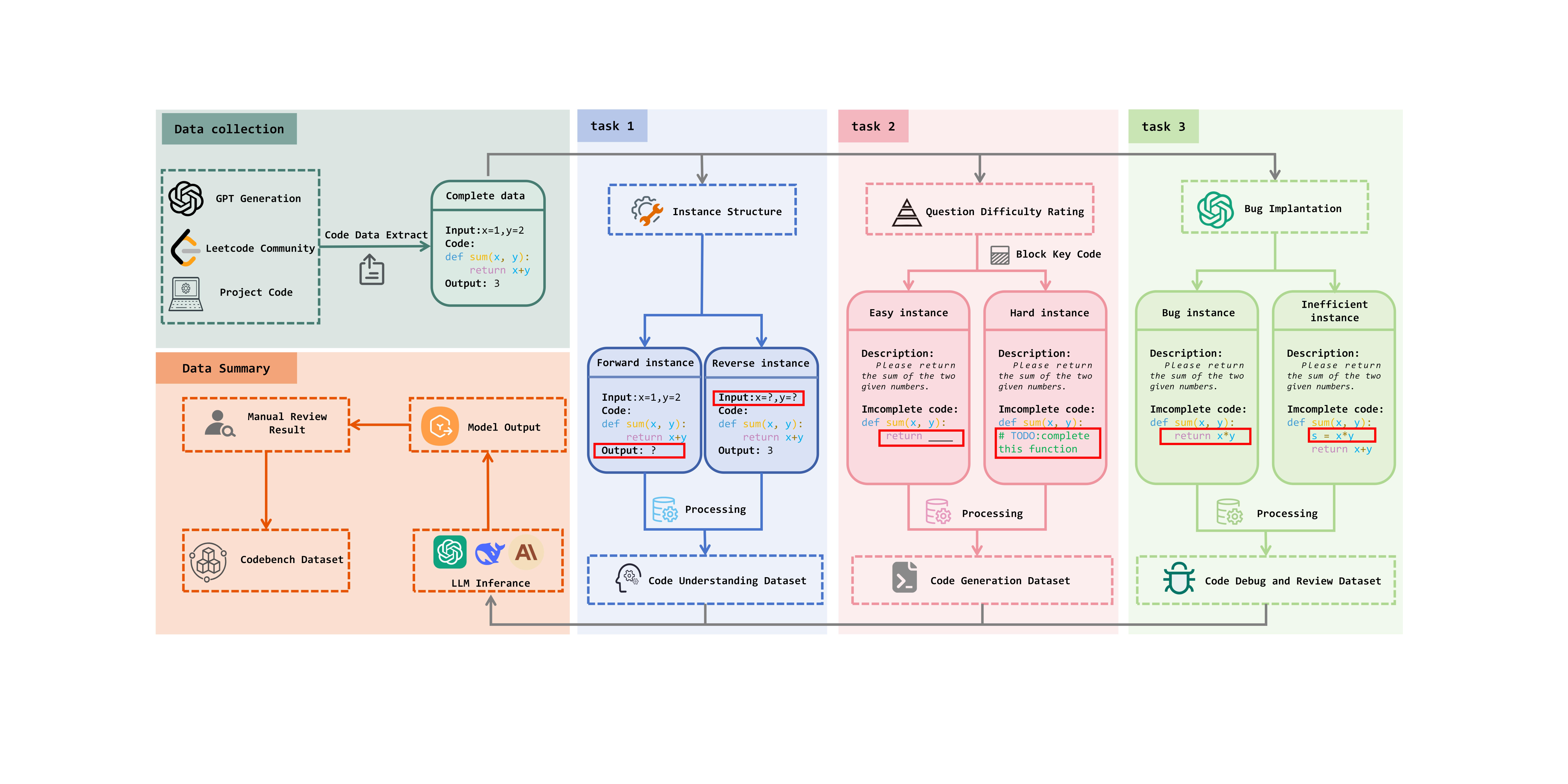} 
\caption{The construction pipeline of CoCo-Bench. The process starts with data collection from three primary sources: GPT generation, Leetcode community solutions, and various projects. The collected data is then structured into samples, which are adapted to different tasks based on their characteristics. \textbf{CU:} the construction of this dataset involves structuring the collected data into forward and reverse instances. \textbf{CG:} this dataset involves the question difficulty rating task, where the collected code is categorized into easy and hard instances. Easy instances are characterized by simple, straightforward tasks with minimal code, whereas hard instances involve more complex tasks with additional comments or incomplete parts that require completion. \textbf{CM} and \textbf{CR} datasets start with a bug implantation process, where code snippets are intentionally altered to introduce bugs and inefficiencies. The final stage of the pipeline involves inference and human check which ensure the correctness and relevance of CoCo-Bench. }
\label{fig3}
\end{figure*}

The sample generation process of CoCo-Bench is strategically organized into three main stages, as shown in Figure~\ref{fig3}: raw data collection, task-specific data transformation and sample review. Each part contributes to the overall effectiveness and quality of the datasets. 

\textbf{Raw Data Collection:} The code collection phase of our dataset construction pipeline is designed to gather a diverse set of programming samples that adhere to strict timeliness criteria, preventing data contamination and ensuring the quality and relevance of the data for training and validating machine learning models. We source code from Leetcode and various project repositories, selected under stringent conditions to ensure data freshness. Leetcode provides a controlled environment with frequently reviewed and updated code, incorporating current coding practices and algorithms. This is crucial for training models that must stay up-to-date with the latest programming trends. Project code, particularly from repositories with contributions from less experienced developers, serves a dual purpose. It captures common coding errors and suboptimal practices, valuable for training models aimed at CR and CM tasks. Additionally, it aligns the dataset with real-world programming tasks, as project code often involves complex, operationally-driven designs. By carefully selecting sources like Leetcode and project repositories, we ensure that our dataset avoids outdated or irrelevant data while encompassing a broad spectrum of real-world coding scenarios.

\textbf{Task-Specific Data Transformation:} The second step in our dataset construction pipeline involves task-specific data transformation, tailoring code samples for different computational tasks to optimize model training and evaluation. For CU, the process creates forward instances (input-output pairs) and reverse instances (output-inference challenges). In CG tasks, instances are categorized by difficulty, ranging from simple completions to complex corrections. For CM and CR tasks, bugs and inefficiencies are strategically inserted into code samples, using models like GPT to generate realistic scenarios, enhancing the dataset’s robustness for debugging and optimization tasks. The final step includes a thorough CR, starting with initial evaluations using high-performance open-source models like GPT, followed by manual review, and final incorporation into the benchmark library. Following the automated inference, the samples undergo a detailed manual review by experienced developers. The manual review focuses on several critical aspects: 

1. \textbf{Sample Correctness}: Verifying that the sample code performs the intended function without errors. This involves running the code against a set of test cases to ensure accurate outputs. 

2. \textbf{Reasonable Difficulty}: Assessing whether the difficulty level of the sample is appropriate. Ensuring that the samples are neither too easy nor excessively difficult, thereby maintaining a balanced difficulty level across the dataset. 

3. \textbf{Practical Applicability}: Evaluating the real-world relevance and usefulness of the sample. Ensuring that the code samples reflect practical scenarios and challenges that developers are likely to encounter. 

4. \textbf{Easy Readability}: Checking if the sample code is easy to understand and maintain. This includes verifying clear and concise variable names, appropriate use of comments, and adherence to coding standards. 

During the manual review, each aspect is meticulously examined to ensure that the benchmark samples meet high standards of quality and robustness. 

%% file: src/appendix/4_benchmark_metric.tex
\section{Benchmark Metric} \label{app:metrics}
To evaluate the performance of Code LLMs on the CoCo-Bench, we incorporate a weighting scheme based on the difficulty of each sample, ensuring a more accurate assessment of a model’s capabilities.

We first apply TrueSkill2~\citep{minka2018trueskill} approach to assess the difficulty coefficient of each task, marked as \(\mu_i\). We then calculate \( w_i \) using inverse normalization:
\[
w_i = \frac{1/\mu_i}{\sum_{j=1}^{5} 1/\mu_j}
\]
ensuring \(\sum_{i=1}^{5} w_i = 1\).
Let \( w_i \) represent the weight assigned to the \( i \)-th sample, reflecting the contribution of each task to the overall score. Higher weights correspond to more complex samples. For all tasks, we define the difficulty-aware pass rate (DAPR) as follows: 
\[
\text{DAPR} = \frac{\sum_{i=1}^{n} (\text{Pass Rate}_i \times w_i)}{\sum_{i=1}^{n} w_i}
\]
where \( \text{Pass Rate}_i \) represents the pass rate for the \( i \)-th sample—indicating how often the model successfully passes this specific test case—\( w_i \) is the difficulty weight of the \( i \)-th sample, and \( n \) is the total number of samples in the task.

To compute the overall CoCo-Score, we combine DAPR of each task: 

\begin{align*}
\text{CoCo-Score} = & \sum_{j=1}^{4} \left( \frac{n_j}{N} \times \text{DAPR}_j \right)
\end{align*}
where \( n_j \) is the number of samples in the \( j \)-th task, \( N \) is the total number of samples across all tasks, \( \text{DAPR}_j \) is the difficulty-aware pass rate for each respective task. 

Our approach ensures that the CoCo-Score not only measures the average performance of the models but also emphasizes their ability to handle more complex coding challenges by using difficulty-aware scores, providing a comprehensive assessment of a model's practical effectiveness and robustness across various coding tasks.

%% file: src/appendix/5_prompts_for_inference.tex
\section{Prompts for Inference}
In specific tasks within CoCo-Bench, the input typically consists of a prefix prompt and a suffix prompt. The prefix prompt includes several examples to assist the large model in understanding the specific task requirements and the expected output format. These examples effectively guide the model, ensuring it performs the task correctly. The suffix prompt appears at the end of the input and serves to reinforce the required output structure, ensuring that the model's output is consistent with the examples provided, which facilitates subsequent automated processing. \\

\subsection{Prompts for $\mathrm{CU}$ Inference}
\noindent
\begin{tcolorbox}[colback=bluebackground, colframe=blueframe, title=\textbf{Prefix prompt for \texttt{$\mathrm{CU_{F}}$ }:}, sharp corners, breakable, listing only]
Please deduce the output of the following code based on the code snippet and the input.
\end{tcolorbox}
\noindent
\begin{tcolorbox}[colback=bluebackground, colframe=blueframe, title=\textbf{Infix prompt for \texttt{$\mathrm{CU_{F}}$ }:}, sharp corners, breakable, listing only]
The code snippet is as follows:
\end{tcolorbox}
\noindent
\begin{tcolorbox}[colback=bluebackground, colframe=blueframe, sharp corners, breakable, listing only, boxsep=0pt  
]
\lstset{
    language=Python,
    basicstyle=\ttfamily\fontsize{9}{10}\selectfont,  
    showstringspaces=false,
    frame=none,
    breaklines=true
}
\begin{lstlisting}
import numpy as np
def power_sum(arr1,arr2):
    powered_arr = np.power(arr1,arr2)
    result_sum = np.sum(powered_arr)
    return result_sum
\end{lstlisting}
\end{tcolorbox}

\noindent
\begin{tcolorbox}[colback=bluebackground, colframe=blueframe, title=\textbf{Infix prompt for \texttt{$\mathrm{CU_{F}}$ }:}, sharp corners, breakable, listing only]
The input is as follows:
\end{tcolorbox}
\noindent
\begin{tcolorbox}[colback=bluebackground, colframe=blueframe, sharp corners, breakable, listing only, boxsep=0pt  
]
\lstset{
    language=Python,
    basicstyle=\ttfamily\fontsize{9}{10}\selectfont,  
    showstringspaces=false,
    frame=none,
    breaklines=true
}
\begin{lstlisting}
[[2,3,4], [1,2,3]]
\end{lstlisting}
\end{tcolorbox}

\noindent
\begin{tcolorbox}[colback=bluebackground, colframe=blueframe, title=\textbf{Suffix prompt for \texttt{$\mathrm{CU_{F}}$ }:}, sharp corners, breakable, listing only]
Give only the deduced output of the code snippet. Do not output any additional information.
\end{tcolorbox}
\vspace{0.5cm}

\noindent
\begin{tcolorbox}[colback=bluebackground, colframe=blueframe, title=\textbf{Prefix prompt for \texttt{$\mathrm{CU_{R}}$ }:}, sharp corners, breakable, listing only]
Please deduce the input of the following code based on the code snippet and the output.
\end{tcolorbox}

\noindent
\begin{tcolorbox}[colback=bluebackground, colframe=blueframe, title=\textbf{Infix prompt for \texttt{$\mathrm{CU_{R}}$ }:}, sharp corners, breakable, listing only]
The code snippet is as follows:
\end{tcolorbox}
\noindent
\begin{tcolorbox}[colback=bluebackground, colframe=blueframe, sharp corners, breakable, listing only, boxsep=0pt  
]
\lstset{
    language=Python,
    basicstyle=\ttfamily\fontsize{9}{10}\selectfont,  
    showstringspaces=false,
    frame=none,
    breaklines=true
}
\begin{lstlisting}
import numpy as np
def power_sum(arr1,arr2):
    powered_arr = np.power(arr1,arr2)
    result_sum = np.sum(powered_arr)
    return result_sum
\end{lstlisting}
\end{tcolorbox}

\noindent
\begin{tcolorbox}[colback=bluebackground, colframe=blueframe, title=\textbf{Infix prompt for \texttt{$\mathrm{CU_{R}}$ }:}, sharp corners, breakable, listing only]
The output is as follows:
\end{tcolorbox}
\noindent
\begin{tcolorbox}[colback=bluebackground, colframe=blueframe, sharp corners, breakable, listing only, boxsep=0pt  
]
\lstset{
    language=Python,
    basicstyle=\ttfamily\fontsize{9}{10}\selectfont,  
    showstringspaces=false,
    frame=none,
    breaklines=true
}
\begin{lstlisting}
102
\end{lstlisting}
\end{tcolorbox}
 
\noindent
\begin{tcolorbox}[colback=bluebackground, colframe=blueframe, title=\textbf{Suffix prompt for \texttt{$\mathrm{CU_{R}}$ }:}, sharp corners, breakable, listing only]
Give only the deduced input of the code snippet. Do not output any additional information.
\end{tcolorbox}

\subsection{Prompts for $\mathrm{CG}$ Inference}
\begin{tcolorbox}[colback=bluebackground, colframe=blueframe, title=\textbf{Prefix prompt for \texttt{$\mathrm{CG}$ }:}, sharp corners, breakable, listing only]
Please fill in the following incomplete code according to the description. The description is as follows:
\end{tcolorbox}
\noindent
\begin{tcolorbox}[colback=bluebackground, colframe=blueframe, sharp corners, breakable, listing only]
You are given an array of positive integers nums. Alice and Bob are playing a game. In the game, Alice can choose either all single-digit numbers or all double-digit numbers from nums, and the rest of the numbers are given to Bob. Alice wins if the sum of her numbers is strictly greater than the sum of Bob's numbers. Return true if Alice can win this game, otherwise, return false.
\end{tcolorbox}

\begin{tcolorbox}[colback=bluebackground, colframe=blueframe, title=\textbf{Infix prompt for \texttt{$\mathrm{CG}$ }:}, sharp corners, breakable, listing only]
The incomplete code is as follows:
\end{tcolorbox}
\noindent
\begin{tcolorbox}[colback=bluebackground, colframe=blueframe, sharp corners, breakable, listing only, boxsep=0pt,  
]
\lstset{
    language=Python,
    basicstyle=\ttfamily\fontsize{9}{10}\selectfont,  
    showstringspaces=false,
    frame=none,
    breaklines=true
}
\begin{lstlisting}
def canAliceWin(self, nums: List[int]) -> bool:
    single=0
    double=0
    for it in nums:
        if it>=10:
            double=____
        else:
            single=____
    return single !=double
\end{lstlisting}
\end{tcolorbox}

\begin{tcolorbox}[colback=bluebackground, colframe=blueframe, title=\textbf{Suffix prompt for \texttt{$\mathrm{CG}$ }:}, sharp corners, breakable, listing only]
Give only the completed code. Do not output any additional information.
\end{tcolorbox}

\subsection{Prompts for $\mathrm{CM}$ Inference}
\noindent
\begin{tcolorbox}[colback=bluebackground, colframe=blueframe, title=\textbf{Preffix prompt for \texttt{$\mathrm{CM}$}:}, sharp corners, breakable, listing only]
Please correct the following code according to the description. The description is as follows:
\end{tcolorbox}
\noindent
\begin{tcolorbox}[colback=bluebackground, colframe=blueframe, sharp corners, breakable, listing only]
You are given a 0-indexed string s typed by a user. Changing a key is defined as using a key different from the last used key. For example, s = "ab" has a change of a key while s = "bBBb" does not have any. Return the number of times the user had to change the key. Note: Modifiers like shift or caps lock won't be counted in changing the key that is if a user typed the letter 'a' and then the letter 'A' then it will not be considered as a changing of key.
\end{tcolorbox}

\begin{tcolorbox}[colback=bluebackground, colframe=blueframe, title=\textbf{Infix prompt for \texttt{$\mathrm{CM}$ }:}, sharp corners, breakable, listing only]
The code to be corrected is as follows:
\end{tcolorbox}
\noindent
\begin{tcolorbox}[colback=bluebackground, colframe=blueframe, sharp corners, breakable, listing only, boxsep=0pt,  
]
\lstset{
    language=Python,
    basicstyle=\ttfamily\fontsize{9}{10}\selectfont,  
    showstringspaces=false,
    frame=none,
    breaklines=true
}
\begin{lstlisting}
def countKeyChanges(self, s: str) -> int:
    if len(s) == 1:
        return 0
    s = s.upper()
    count = 0
    for i in range(len(s)-1):
        if s[i] == s[i + 1]:
            count += 1
    return count
\end{lstlisting}
\end{tcolorbox}

\noindent
\begin{tcolorbox}[colback=bluebackground, colframe=blueframe, title=\textbf{Suffix prompt for \texttt{$\mathrm{CM}$}:}, sharp corners, breakable, listing only]
Give only the corrected code. Do not output any additional information.
\end{tcolorbox}

\subsection{Prompts for $\mathrm{CR}$ Inference} 
\noindent
\begin{tcolorbox}[colback=bluebackground, colframe=blueframe, title=\textbf{Prefix prompt for \texttt{$\mathrm{CR}$}:}, sharp corners, breakable, listing only]
Please find errors in the following code according to the description. The description is as follows:
\end{tcolorbox}
\noindent
\begin{tcolorbox}[colback=bluebackground, colframe=blueframe, sharp corners, breakable, listing only]
Function uses the 'eval' function to execute dynamic expressions from user inputs, posing serious security risks.
\end{tcolorbox}

\begin{tcolorbox}[colback=bluebackground, colframe=blueframe, title=\textbf{Infix prompt for \texttt{$\mathrm{CR}$ }:}, sharp corners, breakable, listing only]
The code with errors is as follows:
\end{tcolorbox}
\noindent
\begin{tcolorbox}[colback=bluebackground, colframe=blueframe, sharp corners, breakable, listing only, boxsep=0pt,  
]
\lstset{
    language=Python,
    basicstyle=\ttfamily\fontsize{9}{10}\selectfont,  
    showstringspaces=false,
    frame=none,
    breaklines=true
}
\begin{lstlisting}
def execute_expression(user_input):
    result = eval(user_input) # Dangerous use of eval
    return result
\end{lstlisting}
\end{tcolorbox}

\begin{tcolorbox}[colback=bluebackground, colframe=blueframe, title=\textbf{Suffix prompt for \texttt{$\mathrm{CR}$}:}, sharp corners, breakable]
There are four types of errors: performance\_issues, security\_issues, syntax\_errors and logical\_errors. Please give accurate error types and correct the code, in the form of\\
\{\\
    \hspace*{2em}"performance\_issues":\\
    \hspace*{4em}"data = request.get(user\_url)",\\
    \hspace*{2em}"security\_issues":\\
    \hspace*{4em}"password = getpass.getpass()",\\
    \hspace*{2em}"syntax\_errors":\\
    \hspace*{4em}"print(a + b)",\\
    \hspace*{2em}"logical\_errors":\\
    \hspace*{4em}"continue if a > b else break"\\
\}
\end{tcolorbox}

%% file: src/appendix/6_more_hyper_parameter_analysis.tex
\section{More Hyper Parameter Analysis}

\begin{figure}[h]
    \centering
    \includegraphics[width=0.8\linewidth]{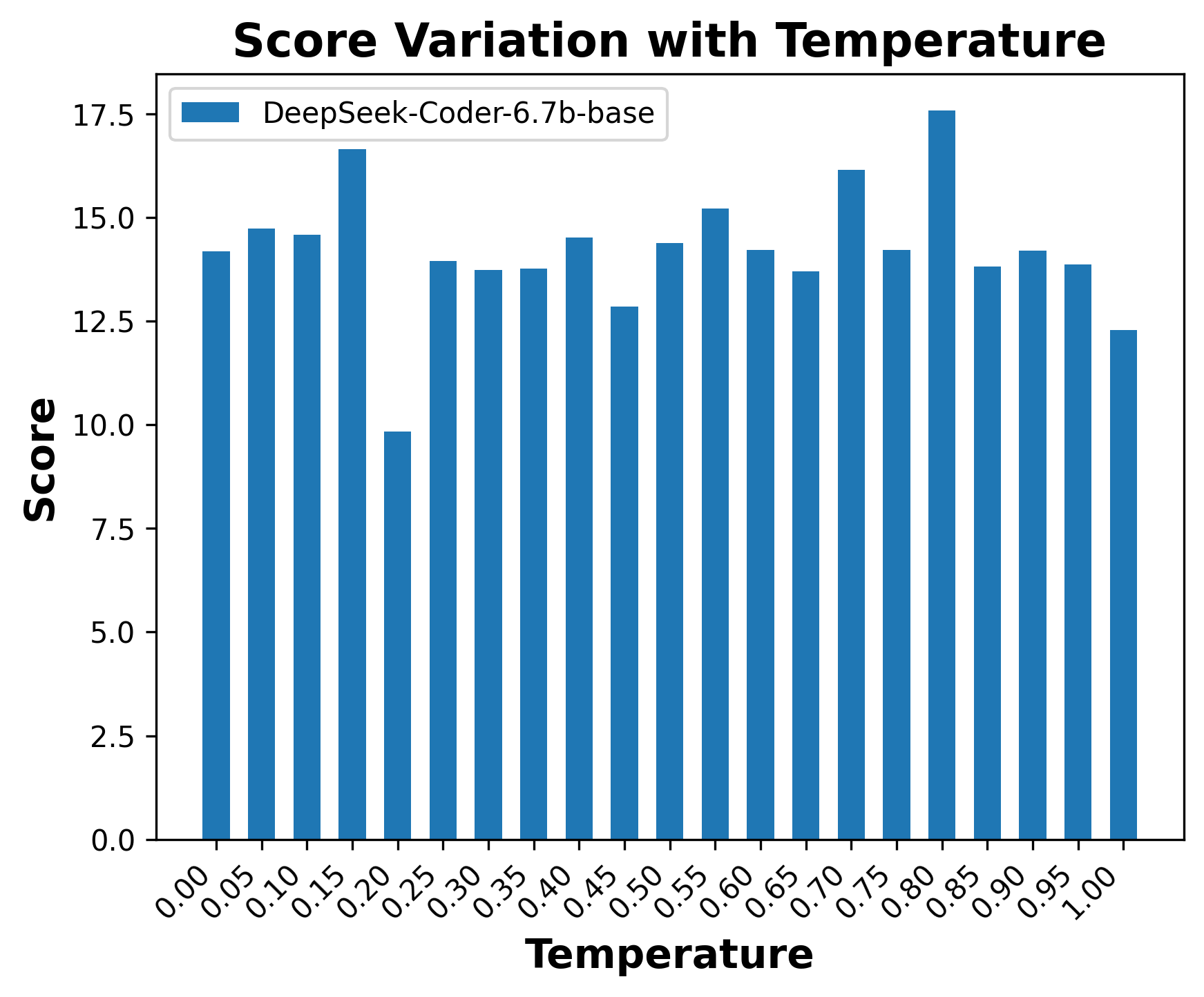}
    \caption{CoCo-Score fluctuations across different decoding temperatures for DeepSeek-Coder-6.7b-Instruct.}
    \label{fig:score_3}
\end{figure}
\begin{figure}[h]
    \centering
    \includegraphics[width=0.8\linewidth]{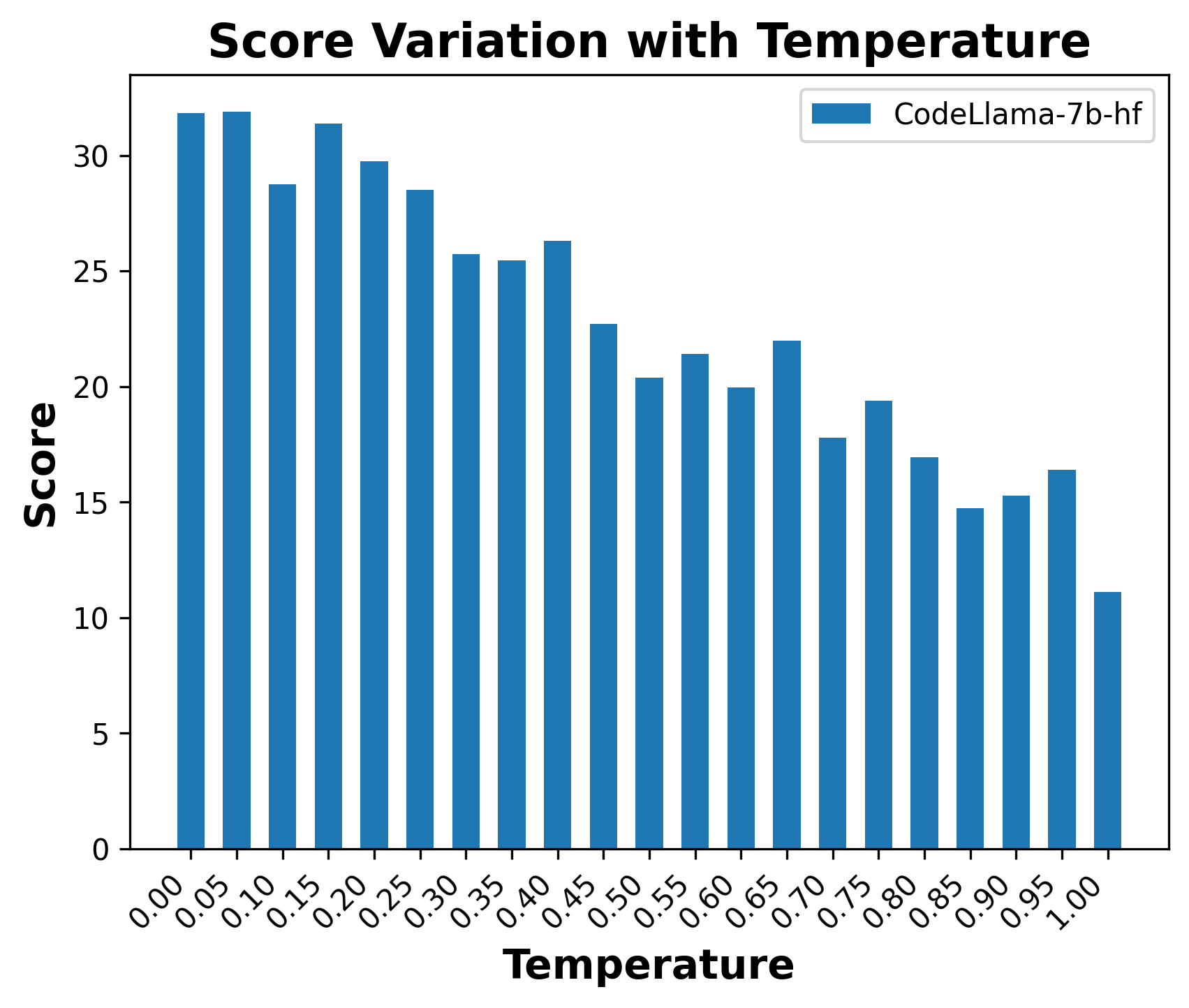}
    \caption{CoCo-Score fluctuations across different decoding temperatures for CodeLlama-7b-hf.}
    \label{fig:score_4}
\end{figure}

An interesting phenomenon can be observed from these Figure~\ref{fig:score_3} and Figure~\ref{fig:score_4}: models with lower performance seem to be more sensitive to decoding temperature. Specifically, CodeLlama-7b-hf achieves its highest score at a lower temperature, while DeepSeek-Coder-6.7b-Instruct shows greater score fluctuations at higher temperatures.

This can be explained by the relationship between a model's generation capability and the decoding temperature. Higher-performing models typically have stronger generation capabilities, allowing them to maintain more stable performance across a wider range of temperatures. In contrast, lower-performing models may rely on specific decoding temperatures to enhance the quality of their outputs when tackling complex tasks. For instance, CodeLlama-7b-hf might produce more deterministic outputs at lower temperatures, avoiding the randomness introduced at higher temperatures, which leads to better scores in certain tasks.

On the other hand, lower-performing models are more sensitive to temperature changes, possibly because they struggle to maintain coherence and quality in their outputs at higher temperatures. As the temperature increases, the generated text may become more random, leading to a decline in task performance. This also explains why CodeLlama-7b-hf achieves its highest score at a lower temperature: at lower temperatures, the model produces more deterministic and consistent content, avoiding the noise introduced by unnecessary randomness.